\documentstyle[aps,multicol,epsfig]{revtex}

\def\br{ {\bf r} }

\def\bk{ {\bf k} }
\def\bv{ {\bf v} }
\def\bps{ {\bf p}_s }

\def\im{ \,{\rm Im}\, }
\def\re{ \,{\rm Re}\, }

\begin{document}
\bibliographystyle{unsrt}
\draft

\title{Localized surface states in HTSC: alternative mechanism of
zero-bias conductance peaks}

\author{K.~V.~Samokhin$^*$ and M.~B.~Walker}

\address{Department of Physics, University of Toronto, Toronto, Ontario,
 Canada M5S 1A7}

\date{\today}

\maketitle

\begin{abstract}
It is shown that the quasiparticle states localized in the vicinity of
surface imperfections of atomic size can be responsible for the zero-bias
tunneling conductance peaks in high-$T_c$ superconductors. The contribution
from these states can be easily separated from other mechanisms using
their qualitatively different response on an external magnetic field.
\end{abstract}


\begin{multicols}{2}\narrowtext

One of the striking features of high-$T_c$ superconductors (HTSC) is
the presence of zero-bias peaks in the voltage dependence of
the tunneling conductance (ZBCP's).
These peaks, which have been observed in numerous experiments using
low-temperature scanning tunneling microscopy (STM) technique
\cite{Kash95,Wei98}, planar SIN junctions \cite{Coving},
and grain boundaries Josephson junctions \cite{Alff98},
are generally considered as a clear signature of the unconventional
pairing symmetry in cuprates. Thus it is important to understand what
physical mechanisms can lead to the formation of ZBCP's and how
the contributions from different mechanisms can be separated in experiments.

The most popular model attributes the origin of ZBCP's to the Andreev
surface bound states (ABS's) \cite{ABS}, whose existence in
unconventional superconductors is related to the fact that quasiparticles
reflected from the interface see a change in the sign of the order parameter
along their classical trajectories. The interplay of multiple Andreev and
specular reflections then leads to the formation of zero-energy bound states in
the vicinity of the interface. The theory predicts that the ABS contribution to
ZBCP's is strongly anisotropic reflecting the underlying symmetry of the order
parameter and, in particular, is absent for those orientations of the interface
for which the gap does not change its sign along the quasiparticle's
trajectory. This result, however, contradicts some of the experimental data, in
which no significant dependence of the ZBCP magnitude on the interface
orientation in $a$-$b$ plane has been found \cite{Kash95,Wei98,Coving}.
On the other hand, the analysis of the behaviour of ABS's in an external
magnetic field shows that ZBCP's should split symmetrically, the splitting being
linear in $H$ \cite{Fogel97}. The experimental situation, however, does not lend
unambiguous support to this prediction. In particular, in Refs.
\cite{Kash95,Alff98}, a suppression and broadening of the in-plane ZBCP's has been
observed (see also Ref. \cite{Ekin97}, where similar results for the $c$-axis
tunneling were reported).
 The most plausible explanation of the presence of ZBCP's for all
surface orientations is that the surfaces of real
samples are not perfectly flat at the atomic scale, so the incident
quasiparticles are not reflected specularly, but rather get scattered in all
directions resulting in the formation of ABS, albeit with a smaller spectral
weight \cite{Fogel97,rough}. However, the absence of ZBCP splitting in magnetic
field cannot be explained in the framework of existing theories.

In this article, we propose a new mechanism for the formation of zero-bias anomalies
in $d$-wave superconductors which does not rely on the existence of ABS's.
Briefly, our idea is that a significant contribution to ZBCP's, at least for some
in-plane surface orientations, can come from the states which are localized near atomic-scale
surface imperfections, and thus are very different from ABS's which propagate along
the surface. In contrast to the previous approaches,
we use an essentially non-quasiclassical way of modeling the surface
roughness, namely, we assume that there are strong defects at the surface of a lattice
superconductor, such as missing atoms.
It is known that a single scalar impurity has a notable effect on the bulk
$d$-wave superconducting state, creating an impurity bound state (IBS)
in its vicinity, whose energy and width tend to zero in the limit of strong
impurity potential \cite{IBS}.
Experimentally, IBS's manifest themselves in the existence of sharp zero-bias
peaks in the voltage dependence of the differential tunneling conductance,
which have been observed recently in beautiful STM
experiments on BSCCO compounds \cite{Pan00}.
We will show that ZBCP's in HTSC can be attributed
to the formation of zero-energy IBS's by strong surface defects, similar to
those in the bulk. These states possess the two desirable features: (i) they exist
for the ``anti-node'' surface orientations, and (ii), in contrast to the Andreev
states which split in magnetic field, the IBS peaks remain centered around the zero
energy, but get suppressed and broadened.

\begin{figure}
\begin{center}
\leavevmode \epsfxsize=0.4\linewidth \epsfbox{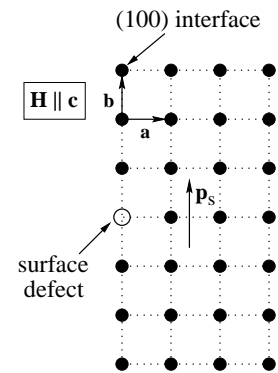}
\caption{Lattice model for a surface defect in a $d$-wave superconductor
in the presence of magnetic field.}
\label{Fig1}
\end{center}
\end{figure}

Let us consider a single missing atom or a strong repulsive point-like impurity
at a (100) surface of a two-dimensional $d$-wave superconductor. (We consider
this geometry because it is the simplest case with no ABS's, see the
discussion in the end of the article.) Because of the inherent
non-locality of the order parameter in a superconductor with higher angular momentum
pairing and the shortness of the coherence
length in HTSC, which is typically of the order of $1nm$, it is convenient to use
the lattice representation of the Hamiltonian:
\begin{equation}
{\cal H}=\sum\limits_{\langle\br\br'\rangle}C^\dagger(\br)
 H(\br,\br')C(\br'),
\end{equation}
where $C(\br)=(c_\uparrow(\br),c^\dagger_\downarrow(\br))^T$
are Nambu operators, and
\begin{equation}
\label{Hbdg}
H=\left(\begin{array}{cc}
H_0 & \Delta \\
\Delta^* & -H_0^*
\end{array}\right).
\end{equation}
is the Bogoliubov-de Gennes (BdG) operator with $H_0(\br,\br')=
-t(\br,\br')+U(\br)\delta_{\br\br'}+\mu\delta_{\br\br'}$. The
lattice sites are labeled by $\br$, and
$\langle\br\br'\rangle$ means nearest neighbors.
We choose the gauge in which the order parameter is real, so the
hopping matrix element has the form $t(\br,\br')=te^{i\bps(\br-\br')}$.
The magnetic field is directed along the $c$-axis, and
the sample is in the Meissner state, so the screening supercurrent
decays in the bulk and can be assumed to be constant
in the vicinity of the surface:
$\bps=-(e/c){\bf A}(x=0)=(e/c)H\lambda{\bf b}$,
where $\lambda$ is the London penetration depth.
The impurity scattering is described by the scalar potential
$U(\br)=u\delta_{\br{\bf 0}}$ with $u\to\infty$ (the
unitary limit), which describes a surface irregularity of atomic size.
The mean-field order parameter $\Delta(\br,\br')$,
corresponding to $d_{x^2-y^2}$ symmetry, is equal to
$+\Delta_0$ for $\br=\br'\pm{\bf a}$, and $-\Delta_0$ for $\br=\br'\pm{\bf b}$,
where ${\bf a},{\bf b}$ are unit vectors of the square lattice
(see Fig. \ref{Fig1}). We do not
calculate the order parameter self-consistently, thus
$\Delta_0$ is assumed to be constant.
This assumption is valid only for the (100) surface orientation,
which follows from the boundary condition for the order parameter
in the absence of an impurity \cite{Sam95}.
The situation could be further complicated if a suppression of the order
parameter near the impurity site was to be taken into account. Although
the numerical investigation of the self-consistency equations
shows that such an effect does exist \cite{Hett99}, we neglect it because
it leads only to a renormalization of the effective impurity strength
towards the unitary limit \cite{Shnir99}.

The quantity measured in tunneling experiments is the differential
conductance, which is proportional to the local density of
states (DoS)
\begin{equation}
\label{DoSdef}
 N(\br,\omega)=-\frac{1}{\pi}\im G^R_{11}(\br,\br;\omega),
\end{equation}
where $G^R$ is the retarded Gor'kov-Nambu matrix Green's function.
In the presence of a point-like scalar static impurity,
$G^R=G^R_0+G^R_0TG^R_0$,
where $G^R_0$ is the Green's function of a half-infinite clean superconductor,
and the $T$-matrix
is given by $T(\omega)=u\tau_3\left[1-u g(\omega)\tau_3\right]^{-1}$.
Here $g(\omega)=G^R_0({\bf 0},{\bf 0};\omega)$, and $\tau_i$ are the
Pauli matrices.
The poles of the $T$-matrix correspond to the energies of the
impurity-induced quasiparticle bound states.
A surface vacancy (missing atom) prevents electrons from residing
at its site and thus can be thought of
as an infinitely strong repulsive impurity, which corresponds to
$u\to\infty$. In this limit, the $T$-matrix
has a compact form $T(\omega)=-g^{-1}(\omega)$, so the local DoS
induced by a surface defect is given by
\begin{equation}
\label{Nr}
 \delta N({\bf r},\omega)=\frac{1}{\pi}\,{\rm Im}
 \left[G_0^R(\br,{\bf 0};\omega)g^{-1}
 (\omega)G_0^R({\bf 0},\br;\omega)\right]_{11}.
\end{equation}

The unperturbed Green's function $G^R_0$ can be expressed in terms
of the eigenfunctions $\Psi_\alpha(\br)$ and eigenvalues $E_\alpha$
of the BdG operator (\ref{Hbdg}): $G_0^R(\br_1,\br_2;\omega)=
\sum_\alpha\Psi_\alpha(\br_1)\Psi^\dagger_\alpha(\br_2)/(\omega_+-E_\alpha)$,
where $\omega_+=\omega+i0$.
An essential ingredient of our theory which makes it different from
the previous work on IBS's in the bulk, is the necessity to
impose some boundary conditions on the quasiparticle wave functions
at the superconductor-vacuum interface. If the surface coincides with
the $x=0$ plane, then the boundary conditions are
$\Psi_{1,2}(x=-d,y)=0$,
where $d$ is the lattice constant.
It is worth mentioning here that if we used a continuum model, then
the wave function would be required to vanish right at the
interface, i.e. at $x=0$. In this case, a point-like surface impurity
would not have any effect, so a more complicated approach to dealing with
surface roughness would be necessary \cite{quasicl}.
In our theory, using the lattice model allows
one to incorporate both the non-locality of the order parameter and the
surface roughness on the atomic scale in a simple and natural way.
The authors of Ref. \cite{Tanuma98} used a somewhat similar approach in their
numerical investigation of the self-consistent solution of the BdG
equations. Our model differs from that of Ref. \cite{Tanuma98} in several
important aspects: (i) we clarify the physical reason behind the appearance of
ZBCP's, namely the formation of bound states, (ii) our configuration of surface
defects and the way of imposing the boundary conditions at the interface are
different,  and (iii) we study the magnetic field response of a surface IBS,
the problem which has not been addressed before.

The quasiparticle wave functions are considerably modified in the
presence of a surface, and the Green's function satisfying the boundary
conditions takes the form
\begin{eqnarray}
\label{G_0}
G^R_0(\br_1,\br_2;\omega)=4\int_0^{\pi/d}\frac{dk_x}{2\pi}
\int_{-\pi/d}^{\pi/d}\frac{dk_y}{2\pi}\;G^R_0(\bk,\omega)\nonumber\\
\times\sin k_x(x_1+d)\sin k_x(x_2+d)e^{ik_y(y_1-y_2)},
\end{eqnarray}
where
\begin{equation}
G^R_0(\bk,\omega)=\frac{(\omega_+-\bv_\bk\bps)\tau_0+\xi_\bk\tau_3+
\Delta_\bk\tau_1}{(\omega_+-\bv_\bk\bps)^2-\xi^2_\bk-\Delta^2_\bk}.
\end{equation}
is the Green's function of a bulk clean superconductor affected by
a ``Doppler shift'' in the quasiparticle energy,
$\xi_\bk=-2t(\cos k_xd+\cos k_yd)-\mu$
is the normal state excitation spectrum, $\bv_\bk=\nabla_\bk\xi_\bk$ is the
Fermi velocity, and $\Delta_\bk=2\Delta_0(\cos k_xd-\cos k_yd)$
is the momentum-dependent superconducting gap. We use the values
of the parameters typical for YBCO compound: $t=185 meV$,
$\mu=0.51 eV$, $\Delta_0=15 meV$.

The energies of the impurity-induced surface bound states satisfy the equation
$\det g(\omega)=0$, whose solutions can be complex. We are interested in the
case of strong impurity scattering and small supercurrent, so that the relevant
energies are expected to be small compared to the magnitude of the gap. At
$\omega,v_Fp_s\ll\Delta_0$, the momentum integrals are restricted to small
vicinities of the two gap nodes $(k_0,k_0)$ and $(k_0,-k_0)$, where $\cos
k_0d=-\mu/4t$, and $v_F=2\sqrt{2}td\sin k_0d$ is the Fermi velocity at the gap
nodes. Introducing the notation
$z=\omega/\Delta_0$, we have
$g(\omega)=(1/4td^2)F(z)\tau_0$
at complex $\omega$, where
\begin{eqnarray}
\label{F}
F(z)=-i z+\frac{1}{\pi}(z+z_s)\ln(z+z_s)&&\nonumber\\
 +\frac{1}{\pi}(z-z_s)\ln(z-z_s)&&
\end{eqnarray}
with $z_s=v_Fp_s/\sqrt{2}\Delta_0$. Two
logarithmic branch cuts are chosen to go down from $z=\pm z_s$ parallel
to the negative imaginary axis.
The equation for the spectrum of bound states in the unitary
limit takes the form $F(z)=0$.
In the absence of supercurrent, $F(z)\to F_0(z)=(2/\pi)z\ln z-i z$, and
the above equation can be easily solved, the solution  being $z_0=0$.
Both the real and imaginary parts of the bare IBS energy vanish, which means
that there is a zero-energy bound state in the vicinity of
a surface vacancy. The presence of this state gives rise to a
sharp peak in the DoS, in full analogy to the
situation in the bulk \cite{IBS}.
At nonzero magnetic field, the dominant energy scale
in the unitary limit is provided by the Doppler shift $v_Fp_s$. It can
be checked that the spectral equation with $F(z)$ given by (\ref{F})
has only one solution in the
complex plane: $z_0=-i\pi z_s/2|\ln z_s|$,
so the IBS energy $\omega_0=z_0\Delta_0$ has the form
\begin{equation}
\label{omega0}
 \re \omega_0=0,\quad
 \im \omega_0=-\frac{\pi v_Fp_s}{2\sqrt{2}}
 \left|\ln\frac{v_Fp_s}{\sqrt{2}\Delta_0}\right|^{-1}.
\end{equation}
From Eqs. (\ref{omega0}) we see that IBS in the unitary limit is
destroyed by supercurrent and replaced by a resonance peak centred around zero
energy, whose width depends non-analytically on $H$, proportional
to $H(\ln H)^{-1}$.
The physical reason for this is clear from Eq. (\ref{DoSdef}): at $p_s\neq 0$,
the bulk DoS does not vanish at $\omega=0$, being proportional to $v_Fp_s$,
which leads to a much stronger hybridization between IBS and the bulk states.
It can be shown that at large but finite impurity strength, the surface
IBS peak is shifted away from zero energy, the shift being proportional to
$H^2$. However, if $v_Fp_s$ is larger than the bare IBS energy \cite{IBS},
the results for the unitary limit are recovered. This behaviour is similar
to what is expected for impurities in the bulk \cite{SW01}.

\begin{figure}
\begin{center}
\leavevmode \epsfxsize=0.9\linewidth \epsfbox{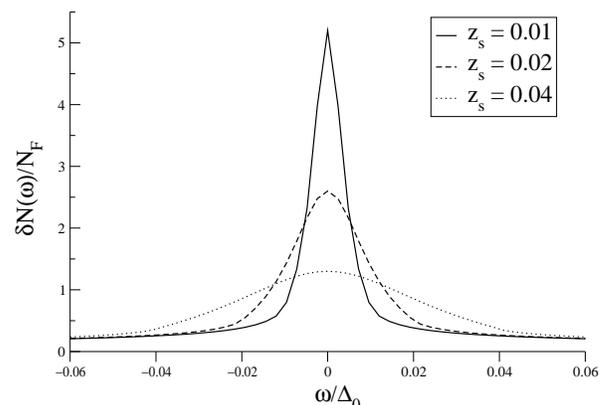}
\caption{The change in the total interface DoS induced by IBS's,
 as a function of energy at increasing supercurrent ($n_i=0.1$).
 A sharp peak at zero field is not shown.}
\label{Fig2}
\end{center}
\end{figure}

In order to visualize our results and facilitate the comparison with
experiments, we compute numerically the surface DoS for a finite concentration
of surface defects. It is easy to see  that
$N_0({\bf 0},\omega)\to 0$ at $u\to\infty$. For this reason, in
order to study the surface IBS, one should calculate either the local
DoS, for example at one of the nearest
neighbors of the impurity site, which can be probed by STM technique,
or the total interface DoS measured in planar tunneling
experiments. Here we concentrate on the latter case, which is
obtained by putting $\br=(0,y)$ in Eq. (\ref{Nr}) followed by the
integration over $y$. The resulting contribution to the total DoS from
the surface IBS's is
\begin{equation}
\label{Ntot}
 \delta N(\omega,p_s)=n_i\im I(\omega)
 F^{-1}\left(\frac{\omega}{\Delta_0}\right),
\end{equation}
where $n_i$ is the linear concentration of surface defects,
$I(\omega)=(4td^2/\pi)\int dy\,[G_0^R(y,\omega)G_0^R(-y,\omega)]_{11}$,
and $F(z)$ is given by Eq. (\ref{F}). We
assume a random distribution of defects and neglect the quantum interference
effects, which gives rise to a prefactor $n_i$ on the right-hand side of Eq.
(\ref{Ntot}) (see the discussion below).
At $(\omega,v_0p_s)/\Delta_0\to 0$, $F^{-1}(\omega)$ is singular,
whereas $I(\omega)$ is not and can be replaced by its value at $\omega=p_s=0$,
which is real. We have plotted the results in Fig.
\ref{Fig2}. The plot confirms that there is a sharp zero-bias
peak due to IBS's in the in-plane tunneling conductance, and the magnetic field
leads to the broadening of this peak near the zero bias, and the
suppression of its magnitude. This behaviour is in a stark contrast to what is
expected for the Andreev states.
It should be mentioned that the numerical results of Ref. \cite{Zhu00} show
that the contribution of bound states at randomly distributed strong impurities
to ZBCP's in the $c$-axis planar tunneling is either negligible, or leads just to
a finite conductance at zero bias. We have studied a different setup, namely the
in-plane tunneling, and come to opposite conclusions, a possible explanation being
that the surface disorder configuration in our system favors the appearance of
a significant IBS contribution, because all the defects lie at the
same line -- the (100) interface.

We would like to emphasize here that the surface impurity states is not
the only mechanism that can lead to the observation of ZBCP's for a nominal (100)
orientation \cite{Fogel97,rough}. It is the magnetic field response of ZBCP's that
should help determine which mechanism gives the dominant contribution.
As said in the introduction, suppression and/or broadening of the in-plane
ZBCP's without any trace of splitting has been seen in some tunneling experiments
(see e.g. Ref. \cite{Kash95} where the STM results for (100) YBCO films were
reported), which qualitatively agrees with our predictions.
However, the possibility of a quantitative comparison of the experimental data
to our results strongly depends on the details of the sample preparation and the surface
quality. The basic assumption of our model is that the microscopic surface
roughness can be described in terms of strong potential scatterers of atomic
size, which is likely to be the case for flat surfaces with missing
atoms or steps. On the other hand, this assumption is definitely wrong if the
characteristic size of the surface imperfections is larger than the coherence
length $\xi_0$. An ideal experimental test of our model should be performed on
a high-quality (100) surface in a broad range of parallel magnetic fields.

In this article, we concentrated on the zero-energy IBS's which are formed in the vicinity of
(100) surfaces. For other surface orientations, the situation is complicated
by the presence of ABS's, which also have zero energy and should therefore experience a
strong hybridization with IBS's. The result of this interplay is not clear {\em a priori}
and requires a separate investigation, which is beyond the scope of the present
study. Other factors which can potentially threaten our results in the presence
of a finite concentration of defects are the multiple-scattering interference
effects and the self-consistent order parameter variation \cite{Atkins00}.
Although their role certainly deserves further analysis,
one can always assume that if the surface disorder concentration is sufficiently
small, then the characteristic energy scales of our problem, such as the Doppler shift,
can be made greater than those at which the ``dangerous'' effects mentioned above
come into play.

In conclusion, we have proposed a new mechanism of the formation of zero-bias
peaks in HTSC.
We predict that the strong defects at an ``anti-node'' surface
can lead to the creation of zero-energy localized states
in their vicinities, whose properties significantly differ from those of the Andreev surface
states.
These localized states should manifest themselves by the presence of sharp zero-bias peaks
in tunneling experiments, which get suppressed and broadened in
an external magnetic field. \\

We would like to thank Patrick Fournier and John Wei for
stimulating discussions.
This work was supported by the Natural Sciences and Engineering
Research Council of Canada.

\end{multicols}


\end{document}